\documentclass[sigconf]{acmart}

\usepackage{array}
\usepackage{enumitem}

\AtBeginDocument{%
  \providecommand\BibTeX{{%
    \normalfont B\kern-0.5em{\scshape i\kern-0.25em b}\kern-0.8em\TeX}}}

\copyrightyear{2022} 
\acmYear{2022} 
\setcopyright{rightsretained} 
\acmConference[CHI '22 Extended Abstracts]{CHI Conference on Human Factors in Computing Systems Extended Abstracts}{April 29-May 5, 2022}{New Orleans, LA, USA}
\acmBooktitle{CHI Conference on Human Factors in Computing Systems Extended Abstracts (CHI '22 Extended Abstracts), April 29-May 5, 2022, New Orleans, LA, USA}
\acmDOI{10.1145/3491101.3519771}
\acmISBN{978-1-4503-9156-6/22/04}

\begin{document}

\title[ ]{How to Train a (Bad) Algorithmic Caseworker: A Quantitative Deconstruction of Risk Assessments in Child Welfare}

%AUTHORS -------------------------------------
\author{Devansh Saxena}
\affiliation{%
  \institution{Dept. of Computer Science, Marquette University}
  \streetaddress{Cudahy Hall, 1313 W Wisconsin Avenue}
  \city{Milwaukee}
  \state{WI}
  \postcode{53233}
  \country{United States}}
\email{devansh.saxena@marquette.edu}

\author{Charlie Repaci}
\affiliation{%
  \institution{School of Medicine, Washington University in St. Louis}
  %\streetaddress{140 St. George Street}
  \city{Saint Louis}
  \state{Missouri}
  %\postcode{53233}
  \country{United States}}
 \email{c.repaci@wustl.edu}

\author{Melanie Sage}
\affiliation{%
  \institution{School of Social Work, University at Buffalo}
  \streetaddress{}
  \city{Buffalo}
  \state{New York}
  %\postcode{32816}
  \country{United States}}
  \email{msage@buffalo.edu}

\author{Shion Guha}
\affiliation{%
  \institution{Faculty of Information, University of Toronto}
  \streetaddress{140 St. George Street}
  \city{Toronto}
  \state{Ontario}
%   \postcode{53233}
  \country{Canada}}
 \email{shion.guha@utoronto.ca}

\renewcommand{\shortauthors}{Devansh Saxena et al.}

\begin{abstract}
Child welfare (CW) agencies use risk assessment tools as a means to achieve evidence-based, consistent, and unbiased decision-making. These risk assessments act as data collection mechanisms and have been further developed into algorithmic systems in recent years. Moreover, several of these algorithms have reinforced biased theoretical constructs and predictors because of the easy availability of structured assessment data. In this study, we critically examine the Washington Assessment of Risk Model (WARM), a prominent risk assessment tool that has been adopted by over 30 states in the United States and has been repurposed into more complex algorithmic systems. We compared WARM against the narrative coding of casenotes written by caseworkers who used WARM. We found significant discrepancies between the casenotes and WARM data where WARM scores did not not mirror caseworkers' notes about family risk. We provide the SIGCHI community with some initial findings from the quantitative de-construction of a child-welfare risk assessment algorithm.
\end{abstract}




\maketitle

\section{Introduction}
Child welfare (CW) agencies began using risk assessment tools in the 1980s to reduce bias and standardize decision-making in cases of possible child abuse and neglect \cite{gleeson1987implementing, doueck1993decision}. Today, risk assessment tools that were designed to be used in conjunction with caseworkers' clinical decision-making are often fed into more complex algorithms to support decision-making. Algorithmic risk scores are often viewed as more neutral than a worker's clinical impression of risk, which may be biased. However, worker bias may be embedded in algorithms themselves, offering a veneer of standardization that disguises the degree to which algorithmic risk scores still represent potentially faulty risk assessment. In some cases, a worker's opinion of risk may account for most of the variance in whether children are removed from their homes, even when it comes to quantitative tools \cite{enosh2015reasoning, lauritzen2018factors}. For a variety of systemic reasons, child-welfare caseworkers nationally have an average of less than two-years of work experience in their positions \cite{edwards2018characteristics}. Caseworkers vary widely on their impressions of family risk, which is known to be influenced by demographic factors such as race and gender of the worker and the families \cite{dettlaff2011disentangling}. This raises additional concerns about the possibility that the data being embedded in algorithmic models further amplifies bias in risk assessment.

When an allegation of abuse is made at the screening hotline, a frontline worker assesses the case based on prior referrals and family history and makes a determination to screen in (or out) the case for an investigation \cite{brown2019toward}. Caseworkers conducting these investigations (e.g., home visits, interviewing referent, parents, relatives, neighbors) typically record information in two forms: \textbf{1) quantitative risk assessments} and \textbf{2) unstructured narrative casenotes} that are electronically stored in the case record. These risk assessments were designed to provide a consistent assessment of risk based on specific risk factors that are believed to predict future abuse/neglect. One such tool, the Washington Assessment of Risk Model (WARM), is scored at the investigation alongside an outcome of maltreatment (i.e., founded, unfounded, inconclusive). Concurrently, caseworkers also write detailed casenotes that contain more contextual information about the family such as a text summary of abuse/neglect, discussion of major risk factors, and explanations of risk factor ratings on the risk assessment. However, over the past decade, artificial intelligence (AI) research has grown exponentially with significant attention being paid to developing algorithms using quantitative data collected from risk assessments while overlooking casenotes as a data source that contain more contextual signals about case circumstances.

With  significant growth in AI research, SIGCHI researchers have become very engaged in understanding how fair \cite{selbst2019fairness, agarwal2018reductions, dwork2012fairness}, accountable \cite{wieringa2020account, hardt2016equality}, transparent \citep{kizilcec2016much, kemper2019transparent}, and explainable \cite{cheng2019explaining, kulesza2013too} algorithms may be developed using a variety of design methods. More specifically, researchers have been very interested in how algorithmic decision-making is carried out in the public sector \cite{ clancy2022reconciling, holten2020shifting, robertson2021modeling, saxena2021framework, saxena2020human}. In this study, we critically examine the WARM risk assessment and predictors from WARM which have been adopted into newer algorithmic models. Specifically, we investigate the congruence between WARM and qualitative casenotes. We also examine the degree to which subjective variables in WARM (e.g., parents' cooperation with agency) impact caseworkers' decision-making. Therefore, in this study, we ask the following over-arching research questions:
%Research questions
\vspace{0.1cm}
\begin {itemize} [leftmargin=*]
    \item \textbf{RQ1:} \textit{Where do qualitative caseworker narratives align with (or diverge from) quantitative structured decision-making WARM assessments in regard to risk factors?}
    \item \textbf{RQ2:} \textit{How do caseworkers' biases and perceptions of families become embedded into quantitative structured decision-making assessments?}
\end{itemize}

\vspace{0.1cm}

\noindent In addressing these research questions, our specific contributions are as follows:

\begin {itemize} [leftmargin=*]
    \item We explore a quantitative deconstruction of WARM predictors and theoretical constructs using public child-welfare data. Our findings show how caseworkers' biases become embedded in structured risk assessments as well as the core discrepancies between the purpose of risk assessments and how they are currently being used.
    \item Specifically, we find that WARM measures a parents' response to a caseworker intervention as opposed to the efficacy of the interventions themselves. In addition, we find significant divergences between WARM risk ratings and the risks indicated in the caseworker narratives. This suggests a disconnect between quantitative and qualitative accounts of, ostensibly, the same underlying phenomenon. 
\end{itemize}

These initial findings are part of a larger work-in-progress research project on the quantitative deconstruction of algorithms employed within the U.S. Child-Welfare System (CWS). 

\section{Methods}
\textit{\textbf{The Dataset.}} \hspace{0.2cm} The secondary data used in this study comes from a dataset, \textit{Factors that Influence the Decision Not to Substantiate a CPS Referral}, housed at the the National Data Archive on Child Abuse and Neglect (NDACAN). The federally-funded project  sought to assess child-welfare system decision-making in the state of Washington \cite{english2002factors}, specifically, 1) to identify the factors that influence the decision \textit{\textbf{not}} to substantiate (i.e., find neglect/abuse) a CWS referral; and 2) to identify the characteristics of CWS referrals that are more likely to be unsubstantiated compared to those that are substantiated. For that project, researchers independently examined 2000 cases, exploring two sources of data: \textbf{1) qualitative casenotes} from narrative portions of case records, which were coded and quantified based on narrative risk factors, and \textbf{2) WARM quantitative data} for the same cases. Table 1 depicts one of the coding schemes used specifically for coding physical abuse (as depicted in casenotes) into numeric data. Please refer to English et al. \cite{english2002factors} for more details about the initial study's coding criteria. They conducted multivariate analysis on WARM data to determine predictors related to unsubstantiation of maltreatment in each dataset in order to ascertain protective factors.

\begin{table}
  \small
  \begin{tabular}{c}
     \toprule
     Physical Abuse \\
     \midrule
    Severity 1 = No marks indicated \\
    Severity 2 = Minor marks \\
    Severity 3 = Numerous or non-minor marks \\
    Severity 4 = Emergency room or medical treatment \\
    Severity 5 = Hospitalization for more than 24 hours \\
    Severity 6 = Permanent disability or death \\
  \bottomrule
 \end{tabular}
  \caption{\small Physical abuse to the head/face/neck. Severity is coded on a scale of 1 (low) through 6 (high).}
 \end{table}

\vspace{0.2cm}
\noindent\textit{\textbf{Analysis.}} \hspace{0.2cm}For the purpose of our study we re-examine this dataset. However, instead of treating narrative data and quantitative data as two independent sources of information, we compare across the two to assess how much of the pertinent information is captured by each. As previously noted, risk assessments were designed to be used in conjunction with caseworkers' contextual and clinical judgments \cite{shlonsky2005next, schwalbe2008strengthening}. However, over the past decade, the decision-making latitude has transitioned towards algorithmic decision-making \cite{saxena2020human}. Therefore, motivated by concerns about how public services decision-making has shifted significantly towards the use of quantitative data from risk assessments to recommend/predict an outcome of interest, we compared these information sources to better understand how structured assessments and algorithms might influence risk analysis.  

Therefore, to answer \textbf{RQ1}, we compared the narrative codes against WARM risk factors to assess how much of the pertinent information needed for decision-making is account for by each of these two sources. First, to assess congruence between quantitative (WARM) risk factors and qualitative (case noted) risks, we compared risk categories extracted from each based on the primary researcher's analysis of risk-rating to assess if they were measuring similar items. The shared categories between the two were then used to calculate the degree to which the cases demonstrated incongruities between risk factors identified in casenotes versus the WARM assessment. Overall risk on WARM is depicted on a scale of 0-5 (0=no risk, 1=low risk, 2=moderately low risk, 3=moderate risk, 4=moderately high risk, 5=high risk). Cases that received a 0-2 risk rating receive a \textit{low standard investigation}\footnote{\textbf{Low standard investigations} are defined as a review of prior CPS involvement and collateral contacts to determine if further investigation should occur. They do not require face-to-face contact with the child or caregiver.}, referred to community-based services, and quickly closed. Whereas, cases with risk ratings of 3-5 were assigned a \textit{high standard investigation} \footnote{\textbf{High standard investigations} includes review of prior CPS involvement, collateral contacts, face-to-face interview with child and caretaker, and additional assessments required to determine whether or not abuse/neglect occurred}. This distinction between CW cases, as established by child-welfare in Washington, allowed us to compare across casenotes and WARM data. A case was considered incongruent if the narrative notes explicitly mentioned risk (rated 3-5) but a "No Risk" rating on WARM was entered, and conversely if the narrative notes explicitly mentioned no risk (rated 0-2) but WARM categorized the risk as "Moderate", "Moderately High", or "High" risk. Additionally, we examined the correlations between WARM risk factors to see if there were any surprising or unusual relationships.

For \textbf{RQ2}, we critically examined the data being recorded by WARM itself and how biases may be embedded within this quantitative assessment. Data was randomly organized into an 80/20 training/testing split to perform backward and forward feature selection to mirror the ways this type of data might be used in predictive decision-making. This informed variable selection for building a multinomial logistic regression model (using 10-fold cross validation) to classify overall risk to the child. Variables' significance to the model was quantified and used to further understand whether caseworkers' perceptions of families potentially impacted how risk was measured. 

\section{Results}
\subsection{Discrepancies between Narrative Coding and WARM Risk Factors (RQ1)}

\begin{figure*}
\includegraphics[scale=0.4]{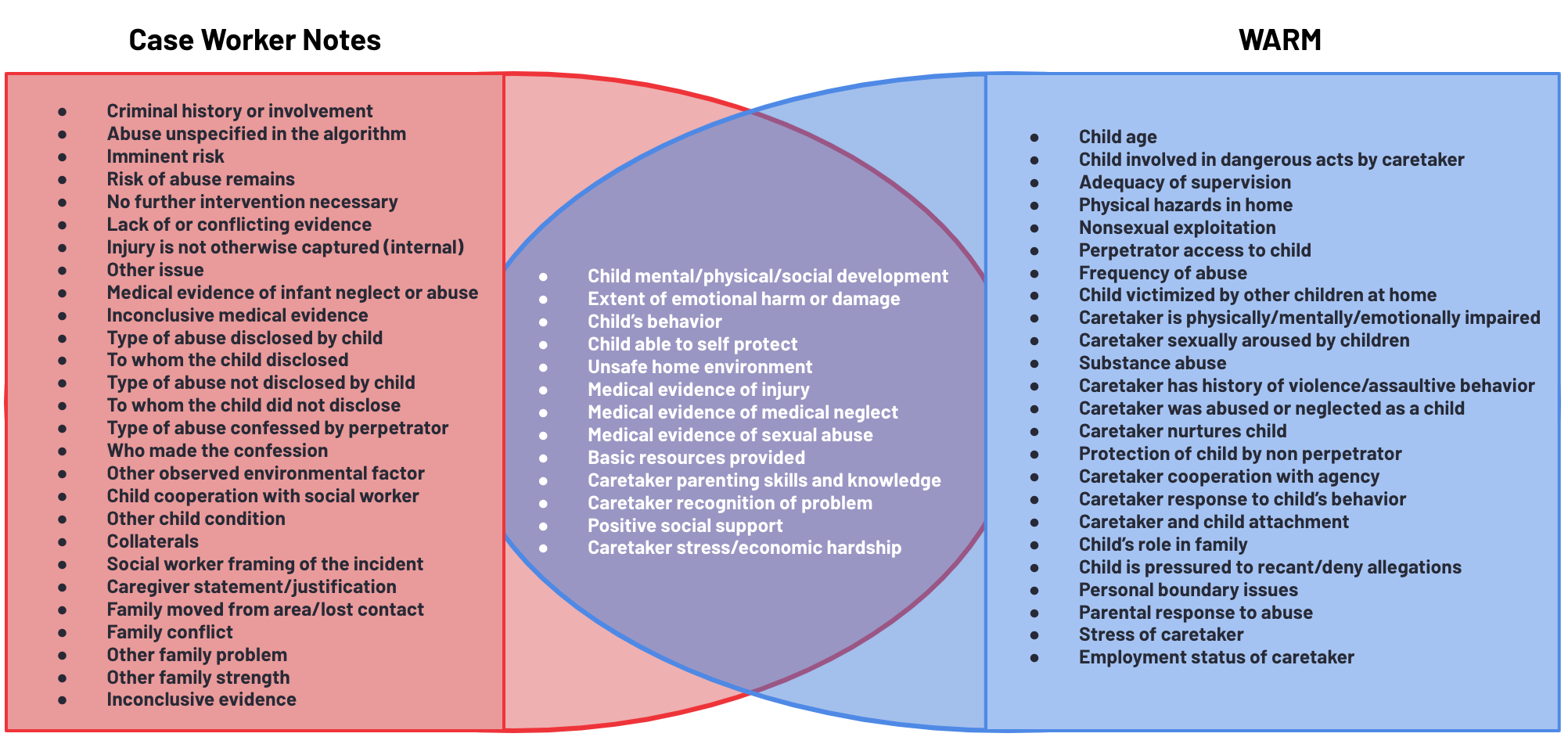}
\caption{Narrative coding variables were compared to WARM categories; while there are some common risk factors, both contain risk factors that the other does not address.}
\end{figure*}

\begin{table}
   \small
   \label{tab:freq}
   \begin{tabular}{lcc}
     \toprule
     Shared Variable&False Positive&False Negative\\
     \midrule
        Child has Bad or Difficult Behavior&7.69\%&13.18\%\\
        Caregiver Doesn’t Recognize Problem&11.25\%&13.07\%\\
        Economic Stress or Hardship&-&27.22\%\\
        Medical Evidence of Medical Neglect&-&3.70\%\\
        Medical Evidence of Physical Injury&-&9.52\%\\
        Medical Evidence of Sexual Abuse&-&25\%\\
        Negative Emotional Condition of Child&8.41\%&17.78\%\\
        No Basic Resources&3.66\%&22.64\%\\
        Positive Social Support&5.23\%&-\\
        Unable to Self Protect&45.35\%&2.30\%\\
        Unsafe Home Environment&2.73\%&31.50\%\\
   \bottomrule
 \end{tabular}
  \vspace{0.2cm}
   \caption{\small WARM to Narrative Coding Comparison. Variables found in both WARM assessment casenotes are used to examine how well WARM works in providing a framework for explaining the case.}
 \end{table}

As depicted in Figure 1, several differences exist between the risk factors mentioned in narrative coding versus the WARM assessment. This could be for several reasons - 1) caseworkers felt constrained by WARM risk factors because they do not appropriately capture safety concerns, 2) different risk/safety priorities emerged during the investigations but were not capture by WARM, 3) depending on when casenotes were completed, the caseworkers' memory might be erroneous or focused on different factors, or overall risk may have shifted over the course of the assessment. However, as depicted in Table 2, caseworkers' casenotes and WARM were aligned in capturing several risk factors, however, there are significant discrepancies in how these factors are scored.

\vspace{0.1cm}
The risk factors common to both the narrative coding and WARM assessment were used to investigate discrepancies in the computed risk. As previously noted, a risk rating of 0-2 results in a \textit{low standard investigation} and a risk rating of 3-5 results in a \textit{high standard investigation}. Therefore, if the casenotes contained a risk rating of 0-2 and WARM assessment for the same case contained a risk rating of 3-5 (or vice versa), a core discrepancy exists. Based on this, incongruities (i.e., \textit{false negatives} and \textit{false positives}) were calculated. False negatives were measured by dividing the number of cases for which casenotes explicitly described risk (rated 3-5) but were marked "No Risk" in the WARM assessment by the total number of cases for which risk was present in casenotes. On the other hand, false positives were calculated by adding all cases marked "Moderate", "Moderately High", and "High" risk in WARM (rated 3-5) where the casenotes explicitly mentioned no risk, and dividing that by the total number of cases for which the narrative coding determined there was no risk present. It is important to note that these percentages are not accuracy measures of WARM itself. The caseworker is the one to both determine the risk to the child according to WARM and also the one to write the casenotes. Instead, these percentages measure where WARM does not align with the content of casenotes. For example, a caseworker may determine that sexual abuse of a child did occur and indicated this on WARM resulting in a classification of high risk. However, they may also determine that due to primary caregiver's protective actions, no ongoing risk (i.e., impending danger) exists within the family, resulting in a classification of no risk in the casenotes. 

\vspace{0.1cm}
This further emphasises the argument that critical information needed for decision-making exists in \textit{\textbf{both}} the casenotes and quantitative WARM data. Casenotes offer more contextual details that highlight the complexities and uncertainties within a case and are necessary for explaining why the WARM assessment contains certain risk ratings as well as the protective factors within a family that may be mediating risk factors. In our recent work, we conducted the first computational inspection of child-welfare casenotes and highlighted how contextual factors such as patterns of invisible labor, impact of systemic factors and constraints, as well as underlying power relationships can be computationally derived from caseworkers' narratives \cite{saxena2022chi}. 

\begin{figure}[]
\includegraphics[scale=0.40]{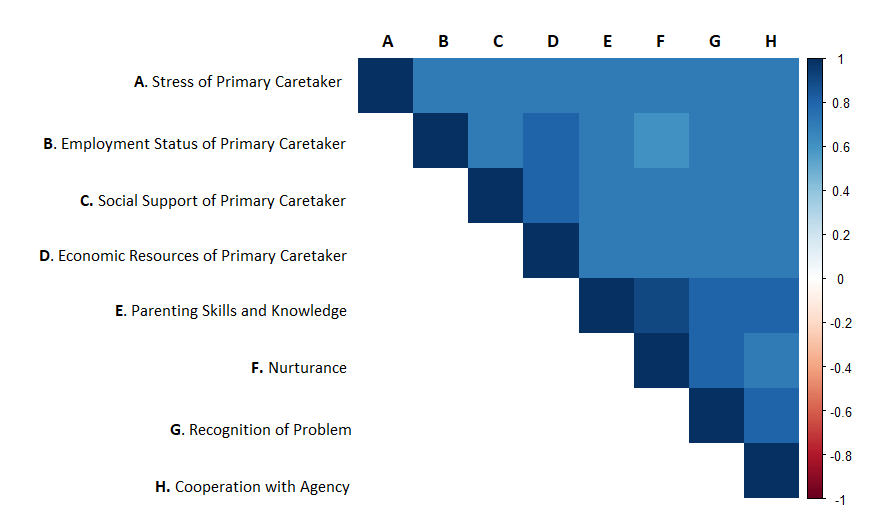}
\caption{Correlations between subjective and related categories regarding caretaker's quality of care.}
\end{figure}

\subsection{Caseworkers' Subjective Impressions are Embedded in Risk Assessments (RQ2)}

Table 3 depicts the most significant variables from WARM that predict the likelihood of maltreatment. These variables were selected via backward step-wise feature selection; the regression model created by forward step-wise feature selection had comparable results. Both models found \textit{Cooperation with agency} and \textit{Stress of caretaker} to be significant predictors of overall risk of recurrence of neglect/abuse. Cooperation with agency additionally has the highest parameter estimate and odds ratio. The above model had a correct negative classification of risk of 86.60\% and a correct positive classification of risk of 80.00\%. As the rating of risk for nurturance, stress of caretaker, age of child, and cooperation with agency increases, the probability that the case is categorized as one with further overall risk to the child also increases. Notably, most of these significant variables are scored by caseworkers and are subjective based upon the caseworkers' impression of the family. 

As depicted in Figure 2, correlations among WARM risk factors were explored in order to assess relationships. Among those that were more expected, such as the high correlation between economic resources, employment status of the caretaker, and the caretaker's stress, there were some very high positive correlations of 0.5 and above between cooperation with agency and other fairly subjective variables related to the quality of care provided by a caretaker. Caretaker's recognition of problem and cooperation with agency were, among others, highly correlated with parenting skills, nurturance, and protection of a child. 

This indicates a potential vector for bias as these predictors are singly scored by caseworkers and are influenced by caseworkers' own impression of the family. As illustrated by prior studies, a caseworker's opinion of risk may account for most of the variance in whether children are removed from their homes, even when it comes to quantitative tools \cite{enosh2015reasoning, lauritzen2018factors, regehr2010confidence}. For example, a caseworker who is struggling to engage with a family may feel that the caretakers are not taking appropriate steps to protect their child by accepting the worker's assistance, or that the caretakers do not have the skills or disposition required to nurture their child due to their rejection of interventions. It is imperative to note here that predictors such as \textit{Parent's Cooperation with Agency} measure parent's response to the intervention rather than the effectiveness of the intervention itself with respect to child safety. In addition, there is a core discrepancy between how the risk assessment is designed to be used versus how the caseworkers are using it. The tool is designed to assess the likelihood of re-referral or recurrence of abuse (i.e., long-term trajectory of risk) \cite{saxena2020human}, however, caseworkers are strictly focused on the substantiation of abuse/neglect at present (i.e., immediate risk posed).

 \begin{table*}[ht!]
  \small
  \label{tab:freq}
  \begin{tabular}{>{\raggedright}p{6cm}>{\raggedright}p{1.3cm}>{\raggedright}p{1.3cm}>{\raggedright}p{2.6cm}>{\raggedright\arraybackslash}p{1cm}}
     \toprule
     Variable & Parameter Estimate & Standard Error & Confidence (95\%) & Odds Ratio\\
     \midrule
    Cooperation with agency\textsuperscript{1}&0.41961 ***&0.085&0.2543 to 0.588&1.5214\\
    Stress of caretaker\textsuperscript{1}&0.34440 ***&0.0841&0.1802 to 0.5103&1.4111\\
    Age of child&0.33551 ***&0.0849&0.1728 to 0.5061&1.3986\\
    Nurturance\textsuperscript{1}&0.39401 **&0.1238&0.1525 to 0.6395&1.4829\\
    Deviant arousal\textsuperscript{1}&-0.57698 **&0.2135&-1.0044 to -0.1671&0.5616\\
    Frequency of abuse/neglect&0.21284 *&0.0845&0.047 to 0.3786&1.2372\\
    Perpetrator access\textsuperscript{2}&0.16153 *&0.0665&0.0301 to 0.2913&1.1753\\
    Attachment and bonding\textsuperscript{2}&0.34011 *&0.1401&0.0681 to 0.6179&1.4051\\
    Child's role in family\textsuperscript{2}&-0.49729 *&0.207&-0.9093 to -0.097&0.6082\\
    History of abuse or neglect as a child\textsuperscript{2}&0.27959 *&0.1268&0.0307 to 0.5289&1.3226\\
    Deviant arousal\textsuperscript{2}&0.31524 *&0.1452&0.0262 to 0.5968&1.3706\\
    Adequacy of medical/dental care& 0.23922 *&0.1172&0.0128 to 0.4741&1.2703\\
  \bottomrule
 \end{tabular}
 \vspace{0.1cm}
 \caption{Significant Variables from Backward Feature Selection. Caretaker\textsuperscript{1} is the primary caregiver (generally biological parent) and Caretaker\textsuperscript{2} is the secondary caregiver in the household.}
 \end{table*}

\section{Discussion}
Algorithms in child-welfare pose a unique set of challenges since interactions between different stakeholders (for e.g., parents and caseworkers) as well as interactions between systemic factors (for e.g., bureaucratic processes and policy-related factors) significantly impact which data is collected about families as well as how decisions are made. In this section, we first discuss the  implications of inconsistencies between casenotes and WARM assessment followed by implications for child-welfare practice. 

\subsection{Inconsistencies between Casenotes and WARM Assessment}
Our results indicate that there are significant inconsistencies between the caseworkers’ casenotes and the WARM assessment scores. Risk assessments were designed to support decision-making and consistently record findings based on a risk framework. That is, they were supposed to be used in conjunction with caseworkers' contextual judgment about cases \cite{schwalbe2008strengthening, shlonsky2005next}. However, structured data collected by these assessments over the past two decades is now  used in algorithms that make high-stakes decisions about children and families \cite{saxena2020human, de2020case}. The lack of internal consistency between caseworker narratives and the WARM assessment discovered in this study is further evidence that algorithmic decision-making likely embeds incomplete assessment data. This further implies that there will be a high degree of uncertainty associated with any predicted outcome and human discretion is needed to fill in these gaps \cite{paakkonen2020bureaucracy, saxena2021framework}. In sum, there are inadequacies in both forms of assessments (caseworker's judgment and algorithmic) such that neither offers a holistic assessment of a family's circumstances. Here, the purpose of this study is to highlight these discrepancies so as to draw attention towards the \textbf{decision-making process} (and how to improve it) instead of the \textbf{decision outcome}. This requires  recognition of the complexities within this socio-political domain, critical decision points, as well as the value-laden choices and heuristic decision-making that CWS staff must engage in. Ongoing engagement with domain experts (for e.g., child-welfare program directors, social work academics, caseworkers, and impacted families) as partners is necessary, rather than a peripheral engagement where computing professionals singly focus on data-driven practices and exercise more agency. Recent work in SIGCHI has highlighted how engagement with public sector stakeholders can uncover the complex decision-making ecosystems \cite{saxena2021framework, ammitzboll2021street}, needs of the stakeholders \cite{cheng2021soliciting, holten2020shifting}, as well as utilize unstructured textual narratives to help contextualize decision-making processes \cite{saxena2022chi}. In addition, the design of research questions, as well as how these questions are situated within the cultural and historical contexts of child-welfare, are of critical importance \cite{redden2020datafied}. Any systems work that is conducted without equal partnerships with child-welfare domain experts will uncritically reproduce and embed similar points of failures (as illustrated by this study) into new sociotechnical systems. 

\subsection{Proxy Variables and Underlying Power Dynamics}
Algorithms are meant to support standardized and evidence-based assessment of risk, thereby reducing bias. However, our results suggest \textit{cooperation with agency} is a significant predictor of risk of maltreatment per the WARM risk assessment. Prior work has established the importance of actively engaging families in interventions and its consequent impact on recurrence of abuse \cite{gladstone2014understanding, toros2018family, smith2008child}. When a tool like WARM quantifies a parent's cooperation with the agency, using it as a proxy for engagement and child safety, worker bias becomes deeply embedded in the algorithm without an adequate critique of the practices that improve (or deteriorate) a family's engagement. That is, risk assessment models often use a family's response to interventions rather than the  effectiveness of the interventions themselves to assess the likelihood of recurrence of abuse. Bureaucratic processes, such as training, protocols, and organizational resources play a significant role in family engagement. For instance, the caseworkers' micro skills, such as collaborative problem solving, focus on strengths, respect for diversity, listening, and reliability have all been presented as worker skills associated with improved engagement \cite{rawlings2019assessing, gladstone2014understanding, smith2008child, toros2018family}. Efforts have been made within child-welfare from both a policy and practice standpoint to transition towards a "Families as Partners" \cite{rauber2009courthouse} model where parents are supposed to act as equal partners in the case planning process and have agency in  decision-making. However, as our results indicate, \textit{cooperation with agency} and \textit{stress of caretaker} are embedded in risk assessment algorithms and scored by caseworkers with no input from parents. That is, a lack of proper examination of variables and underlying biases further shifts power away from parents. In sum, human discretionary work on part of caseworkers and bureaucratic processes significantly impact the context within which (and how) data is collected about families and needs to be critiqued as a critical part of algorithm design. Evidence-based practices in child-welfare service delivery, such as motivational interviewing, describe parent motivation as something that emerges as an outcome of skilled communication led by the caseworker, but motivation is often misplaced as a quality of a parent in risk assessment without an examination of the influence of the caseworker's actions in supporting meaningful engagement \cite{hall2020motivational}.

\subsection{Risk Assessment Algorithms: Old Wine in New Bottles}
Psychometric risk assessments have been used in CWS since the early 1980s, however, there are several imminent concerns in regard to the reliability and predictive validity of variables and outcomes that have now been embedded into and inherited by risk assessment algorithms \cite{gambrill2001need, saxena2020child}. Our results unpack some concerns about one such variable: \textit{cooperation with the agency}. Magura et al. \cite{magura1987assessing} developed the \textit{Family Risk Scales} where the \textit{Parent's Cooperation with Agency} scale originated. Although several predictors from the scale (for e.g., parental recognition of problems, capacity of parents to change, parental motivation etc.) have been repurposed into risk assessment models, there association with recurrence of abuse remains unknown \cite{sunseri2020hidden}. Empirical knowledge related to child-welfare practice is fragmented, and social science theories are needed to fill in these gaps \cite{gambrill2001need}. Computer scientists have continued to focus on the reliability of predictors while averting a closer inspection of predictive validity. A high inter-rater reliability does not necessarily mean that a causal relationship exists between a predictor and an outcome (i.e. - internal validity) \cite{drost2011validity}. As depicted in Table 3, we find that there is a disproportionate effect of specific variables on overall risk. Given that we know how some of those variables (e.g., Stress of Caretaker, Cooperation with Agency etc.) are biased, contextual, and socially constructed, it is no surprise that these power dynamics become embedded in the decision-making outcomes of ascertaining risk. Moreover, a deeper understanding of the impact of interventions (e.g., parenting services), protective factors (e.g., parents' social support system), and the risk posed by the system itself is necessary for developing a more comprehensive understanding of risk \cite{finkelhor2018screening, gambrill2001need}.

\section{Conclusion}
Unpacking a risk assessment algorithm is a good initial step towards understanding the \textit{human discretionary work} and \textit{bureaucratic processes} that influence the decision-making process. Even though such quantitative deconstruction makes visible some of the latent processes that impact the final decision outcome, it still only accounts for a small proportion of caseworkers' day-to-day practices. For instance, in a recent ethnographic study, Saxena et al. \cite{saxena2021framework} found four different risk assessment algorithms that caseworkers used on a daily basis. Moreover, cumulative distrust in algorithms impacted how they interacted with these systems. That is, quantitative deconstruction unveils how caseworker biases and underlying power dynamics can be concealed within structured assessments but a deeper ethnographic analysis of the impact of such tools on caseworkers is equally important. These initial work-in-progress findings are especially important given the veneer of fairness that comes from the increasing use of big data and algorithmic systems.

\begin{acks}
This research was supported by the National Science Foundation grants CRII-1850517 and FAI-1939579. Any opinion, findings, and conclusions or recommendations expressed in this material are those of the authors and do not necessarily reflect the views of our sponsors.
\end{acks}

\bibliographystyle{ACM-Reference-Format}
\bibliography{bibliography}

\end{document}